\documentclass[12pt]{iopart}
\newcommand{\bea}{\begin{eqnarray}}
\newcommand{\beq}{\begin{equation}}
\newcommand{\eea}{\end{eqnarray}}
\newcommand{\eeq}{\end{equation}}

\usepackage[dvips]{graphicx}
\begin{document}

\title[Fluctuations of wave-functions]{Fluctuations of wave functions about 
their classical average}

\author{L Benet 
\footnote[1]{Permanent address: Centro de Ciencias F\'{\i}sicas,
U.N.A.M., Cuernavaca, M\'exico.}, 
J Flores \dag,
H Hern\'andez-Salda\~na \footnote[2]{Instituto de F\'{\i}sica, U.N.A.M., 
Apdo. Postal 20-364, 01000 
M\'exico D.F., M\'exico.}, 
F M Izrailev
\footnote[3]{Permanent address:
Instituto de F\'{\i}sica, B.U.A.P., Apdo. Postal J-48,
72570 Puebla, M\'exico.}, 
F Leyvraz \dag\ and T H Seligman \dag
}
\address{Centro Internacional de Ciencias, Ciudad Universitaria, Chamilpa, 
Cuernavaca, M\'exico.}

\begin{abstract}
Quantum-classical correspondence for the average shape of
eigenfunctions
and the
local spectral density of states are well-known facts.
In this paper, the fluctuations that quantum mechanical wave functions
present
around the classical value are discussed. A simple random matrix model 
leads to a
Gaussian distribution of the amplitudes. 
We compare
this prediction with numerical calculations in chaotic models of coupled
quartic oscillators.
The expectation is broadly confirmed, but
deviations due to scars are observed.
\end{abstract}
\pacs{05.45.-a,05.45.Mt,03.65.Sq}
\maketitle
\section{ Introduction}
At the turn of the century the study of the quantum 
manifestations of classical chaotic systems suffered a significant 
change. Before, the spectral
statistics were amply discussed and it was shown that they follow the 
Random Matrix Theory (RMT) predictions
~\cite{qch,rmt}. The study of 
wave function properties, however, presents 
inherent difficulties
arising from the dependence on the basis used, forcing either to specify one
or  to define basis
independent quantities. Recent progress has been made 
in this respect in the study of  average properties
of eigenstates ~\cite{griba,luna,2body} and of 
the significant statistical deviations from RMT 
~\cite{heller,kaplan}. Nevertheless, not much systematic work exists on the 
eigenfunction fluctuations in dynamical systems ~\cite{griba}. 
In this work we shall contribute on the last subject.
 
In a recent paper \cite{2body} it was found that
the suitably averaged matrix elements between the eigenfunctions (EF)
of two arbitrary Hamiltonians $H_0$ and $H$ are well described
in the semi-classical regime by a classical phase-space integral. 
Specifically, if we define $\phi_\alpha $, $E_\alpha^{(0)}$  to be the 
eigenfunctions and eigenvalues, respectively,
of $H_0$ and $\psi_i$,$E_i$ those of $H$, one finds to a good approximation
\begin{equation}
\bigl\langle\left|\langle\phi_\alpha |\psi_i\rangle\right|^2\bigr\rangle
= \frac{g(E_\alpha^{(0)},E_i)}{\rho_0(E_\alpha^{(0)})},
\label{eq:1}
\end{equation}
where $g(\epsilon, E)$ is given by
\begin{equation}
g(\epsilon,E)= \int dp\,dq\,\delta\left(H_0(p,q)-\epsilon
\right)\delta\left(H(p,q)-E\right),
\label{eq:2}
\end{equation}
which was called in reference \cite{2body} the {\em classical 
eigenfunction} for fixed $E$. 
Here $\rho_0(\epsilon)$ is the level density of $H_0$ calculated by means of 
Weyl's formula. By the symmetry of 
equation (\ref{eq:1}),  the local density of 
states(LDOS)
can be calculated using the energy density of $H$ instead of $\rho_0(\epsilon)$
and maintaining fixed $\epsilon$ in equation (\ref{eq:2}).
For the details, in particular about the way in which the 
l.h.s. of equation (\ref{eq:1}) must be averaged to obtain meaningful results,
see reference ~\cite{2body}. This study was exemplified by two systems of
anharmonic oscillators in one dimension, one coupled
and the other uncoupled. In the present 
letter, our interest is focused on the {\em fluctuations} of
the quantum-mechanical wave functions around this classical limit
in the chaotic case.

\section{A Random-matrix model}
In the usual description of chaotic systems by random matrices, 
the restrictions implied by equations (\ref{eq:1}) and (\ref{eq:2})
are not present. Rather, one attempts to deduce the average 
properties of the eigenfunctions given the structure of the random matrix
ensemble in some particular basis. Instead, here 
we choose pairs of matrices $(H,H_0)$ of size $N\times N$ in such a 
way that the condition
\begin{equation}
\left\langle
\left|\langle\phi_i|\psi_j\rangle\right|^2
\right\rangle
=I_{i,j}
\label{eq:3}
\end{equation}
is always fulfilled, but the matrices are otherwise arbitrary. 
The angular brackets denote the average over the ensemble of matrix pairs
and $I_{i,j}$ stand for any numbers given by outside constraints
such as (\ref{eq:1}). Under these circumstances, we wish to determine 
the full distribution of the matrix elements 
$\langle\phi_i|\psi_j\rangle$. We then proceed to compare 
the predictions of this random matrix model with 
numerical results on models similar to that
studied in \cite{2body}. 

To solve the above problem, we let ourselves be guided by the 
following considerations: the quantities we need to model, 
namely the $\langle\phi_i|\psi_j\rangle$, are nothing else than the matrix
elements of an orthogonal matrix (a unitary one in the case 
where time-reversal invariance is broken, but this does not affect our 
conclusions). We therefore need a random matrix model for
orthogonal matrices with prescribed expectation values for 
the intensities $I_{i,j}$. Note that,
in the large $N$ limit, the Haar measure over the group of orthogonal
matrices can be replaced, up to corrections of order $1/N$, by 
independent Gaussian distributions for all matrix elements $O_{i,j}$,
all having a variance $1/N$. In other words, what we need is  
a random matrix model where the average intensity
$I_{i,j} =\langle|O_{i,j})|^2\rangle$ is given.
If $I_{i,j} \ll 1$, we can consider
$1/I_{i,j}$ as an effective dimension and we expect to get the desired
result, up to corrections of order $O(I_{i,j})$, if we replace the 
average by a simple Gaussian average. We may then, 
to this level of accuracy, take the $O_{i,j}$ as independent
Gaussian variables with variances given by $I_{i,j}$.
 
Note that we postulate a distribution for orthogonal matrices
with the correct values for
$I_{i,j}$. We do not actually derive this distribution, but
simply verify that it has all the required properties. 
If the classical wave function takes very large values and the eigenfunctions
of $H$ do not have a sufficiently large number of components
in the basis of $H_0$, it may happen 
that some $I_{i,j}\simeq1$. Clearly, for these matrix 
elements the model will not apply; we shall see later that 
this happens near peaks or singularities in the classical wave 
function, but then we cannot really make
any comparison with the specific system anyway.
However, for the vast majority of matrix elements, we can expect that
the amplitudes are Gaussian distributed and if we divide the amplitude
$O_{i,j}$ by $\sqrt{I_{i,j}}$ we will find a standard Gaussian.

What deviations from the above predictions should we expect from 
a theoretical point of view?
Clearly, scars \cite{heller} produce an excess of very small amplitudes, 
because
a few exceptionally large amplitudes pick up more of the intensity 
than expected from the classical calculation. Would we also see the large
amplitudes?
Probably not in a statistical analysis against our model, because
these will mainly occur in the region where the classical function
is large and we will usually exclude this region: the condition
$I_{i,j} \ll 1$ is violated there, unless we reach very high
spectral densities, which is scarcely possible in a numerical experiment.
In a real experiment, resolution might well make 
such a high-density region inaccessible also.
If, on the other hand, localization occurs due to disorder or 
due to the fact that the system does not cover the whole phase
space on the Heisenberg time scale \cite{thoulessenergy}, then we may
indeed also see irregularities beyond the realm of very small amplitudes.

However, all deviations mentioned above should only be important if one
of the Hamiltonians, say $H_0$, is integrable in the classical limit.  
If both are chaotic, and we exclude situations in which the two 
Hamiltonians are, in some sense, closely related, we can expect 
not to see any effect of the scars in the amplitudes. The reason for this
can be understood in terms of the traditional picture due to Berry~\cite{beref}
of the  eigenfunctions in phase space: For a chaotic Hamiltonian, they
are expected to cover phase space essentially in a uniform way, up to rather
small concentrations  on periodic orbits. In integrable systems,
on the other hand, eigenfunctions are localized on well-defined tori,
with only half the dimension of the full phase space. The 
overlap between  two chaotic states is therefore far less
likely to become anomalously large than the one between an integrable
state and a chaotic one.

\section{Numerical results}
We now test the Gaussian property against anharmonic oscillator models.
We shall choose the expansion of a chaotic system in terms of an integrable
one; in particular, we have chosen two particles in a quartic
oscillator potential. This ensures a system with scaling
 properties,
for which the classical properties do not change as a function of energy.
We  restrict our attention to antisymmetric wave functions
since in this case we reach the semi-classical limit much more rapidly  than
for the symmetric case.
 The calculation is performed using the
basis of the uncoupled oscillators, which in turn we approximate in a
harmonic oscillator basis \cite{2body}. The Hamiltonian used is
\begin{equation}
H=\sum_{i=1}^n\frac{p_i^2}{2}+\alpha\sum_{i=1}^nx_i^4
+\beta\sum_{1\le i < j}^nx_i^2x_j^2+\gamma
\sum_{1\le i < j}^n\left[
x_ix_j^3+x_i^3x_j
\right],
\label{eq:4}
\end{equation}
where in this case $n$ is equal to two. We have also considered the case $n=4$
with overall similar results \cite{hhst}. We shall use two Hamiltonians: 
one with the same 
parameters as in \cite{2body}, namely $\alpha=10$,
 $\beta=-5.5$ and
$\gamma=5.6$, which we call $H_1$; the other with the 
parameters $\alpha = 10$,  $\beta=\gamma=-4.15$, which
we call $H_2$. The $H_0$ Hamiltonian will have $\alpha=10$ and 
$\beta=\gamma=0$.

\begin{figure}
\includegraphics[width=0.70\textwidth,angle=90,scale=.75]{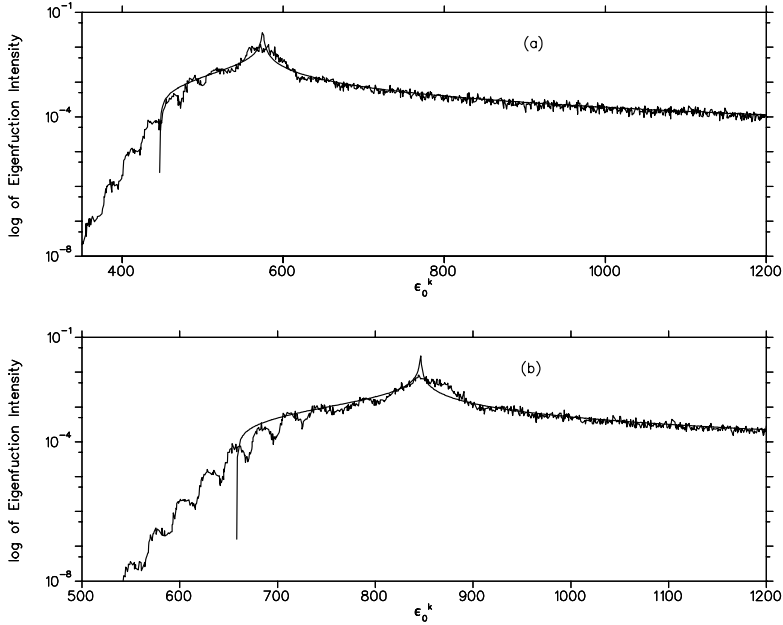}
\caption{\label{efs}Averaged eigenfunction around (a)$\psi_{500}$ and
(b)$\psi_{900}$ 
both for the $H_1$  
Hamiltonian with the parameters given in the text,
with $E_{500}=579.267$ and $E_{900}=846.680$, respectively.
 The smooth curve 
represents the corresponding classical EF. Note the general exponential 
decay plus oscillations in the classically forbidden region.}
\end{figure}

\begin{figure}
\includegraphics[width=0.70\textwidth,angle=90,scale=.75]{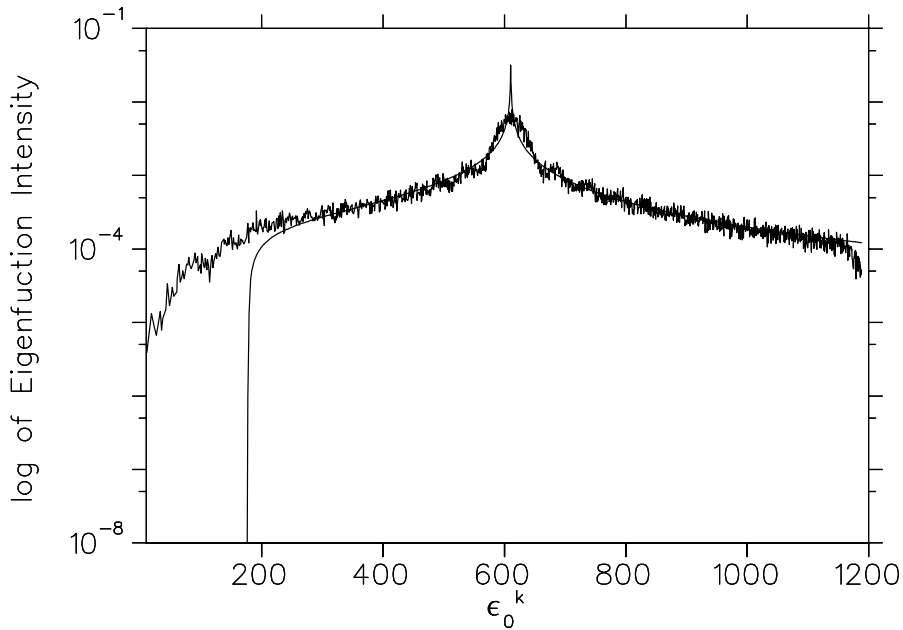}
\caption{\label{efsh2} Averaged eigenfunction around $\psi_{500}$  
for $H_2$ Hamiltonian in the $H_1$ basis with the parameters given in 
the text and 
at energy $E_{500}=611.717$. 
 The smooth curve 
represents the corresponding classical EF.}
\end{figure}

 We analyze the eigenfunctions in terms of the
classical eigenfunction as given in equation (~\ref{eq:2}) at fixed 
energy $E$. The integral is calculated by the Monte Carlo method. 
We find very good
agreement as shown in figures ~\ref{efs} and ~\ref{efsh2}. There the classical EF and 
an average over
101 EF's of the perturbed Hamiltonian are plotted using the method 
of reference ~\cite{2body}.  
Note that the quantum 
functions are not
reliable at the upper end of the classical energy range of $H_0$ for the
higher lying states, although their energies are quite reliable.
At the lower end of the spectra, on the other hand, the amplitudes are very good, and we
find a consistent approximation to an exponential decay of intensities in the
classically forbidden region as shown in figure~\ref{efs}, with 
some system dependent oscillations (these disappear in the 4-particle case 
~\cite{hhst}).

We now proceed to analyze the amplitude fluctuations.  We do this in the 
wings of the
wave functions far from the peak, in regions where the classical function varies slowly
and
is sufficiently small to ensure $I_{i,j} \ll 1$;
of course we restrict our attention to reliable amplitudes. To this end, we first
cut out the parts of the wave function which are either too high in energy 
so that 
they are not reliable, or which lie outside the classically allowed region.
We further cut 4 states on either side of the singularity at the peak of 
the  classical EF. We do this because the fluctuations around the peak 
are large and the peak itself at energy $E$ is a singularity of the 
classical EF. We 
set the norm of the rest to one in both EF and the classical EF. Then we 
proceed to unfold the EF dividing the quantum EF by the classical one, 
defined in equation (~\ref{eq:1}).
In order to compare the different unfolded EFs we renormalize 
them again. 

\begin{figure}
\includegraphics[width=0.70\textwidth,angle=270,scale=.75]{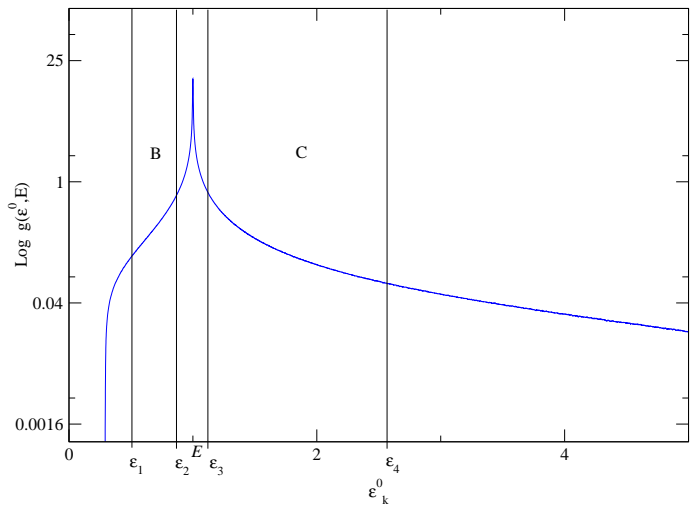}
\caption{ \label{efesq}Eigenfunction tail zones B and C 
considered for
analysis. The extreme cutoffs are determined by the quantum classical 
agreement, and renormalized for unfolding. The center window is $4\Delta$ and 
$80\Delta$ width for the corresponding cases analyzed here.}
\end{figure}

\begin{figure}
\includegraphics[width=0.70\textwidth,angle=90,scale=.75]{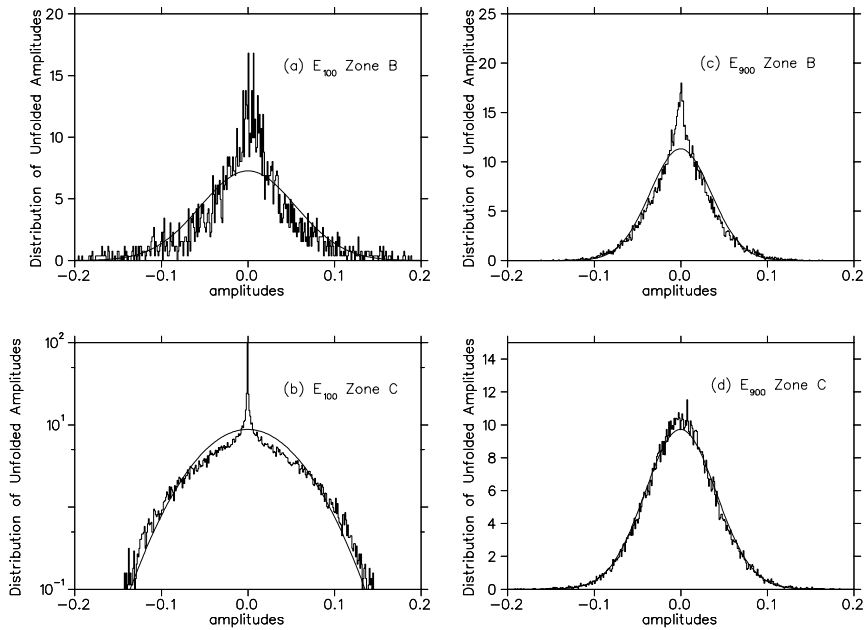}
\caption{ \label{h1dist} Amplitude distribution for the unfolded
EF with energies $E_{100}=204.013 $ (a)-(b) and $E_{900}=846.680$ (c)-(d) 
in zones B and C.
The Gaussian with the same moments as the distributions are indicated by the 
continuous line. Note that (b) is plotted in semilogarithmic scale. }  
\end{figure}

 The corresponding regions in the wings, labeled B and C in
figure ~\ref{efesq}, are the ones with the best quantum-classical 
correspondence. 
We use the intensity 
shape instead of amplitudes for clarity, but all the  calculations were 
performed on the latter.
In order to avoid the rapidly varying region, for the
cases shown below we drop a window of $4$ mean energy level spacings 
$\Delta$ centered in the eigenfunction for $H_1$ and one of $80\Delta$ 
for $H_2$;
the end of the C region is $400\Delta$ away from the center for both 
Hamiltonians (The values of $\Delta$ are, respectively, of $0.826$ and $0.729$.)  
As we cannot perform ensemble averages, we will perform energy
averages within these windows after dividing the amplitudes by the
square root of the local average intensity obtained from the
classical
function, which  agrees well with the quantum average.
As the center of each EF changes in energy, the window center changes but its
width  remains constant.
The amplitude distribution we find, is plotted in figure ~\ref{h1dist}
for the superposition of the
results
of regions B and C on 101 EFs; for low-lying states the shape is far from 
Gaussian 
while for high-lying states we find fair agreement with the Gaussian
behavior except for the excess of small intensities, which we expect
due to scars.
Such scars were seen in the two-body system as exceptional
states
with much narrower intensity distributions and smaller participation
ratios ~\cite{2body}; similar results are found for the four-body 
system ~\cite{hhst}.
Nevertheless, a semi-log plot of the amplitude distribution shows a good 
parabolic 
shape in the wings, even for the zone C in a low-lying states around
$\psi_{100}${}(see figure ~\ref{h1dist}(b)).

\begin{figure}
\includegraphics[width=0.70\textwidth,angle=90,scale=.75]{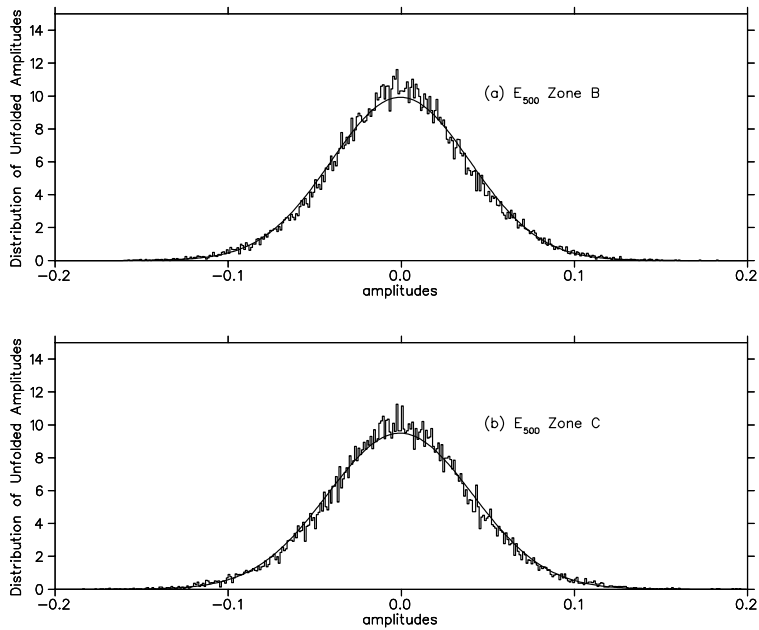}
\caption{\label{h1h2dist} Amplitude distribution for the unfolded
EF $E_{500}=611.717$  of $H_2$ Hamiltonian in the zone B (a)  and C (d). 
The Gaussian with the same moments as the distributions are showed by the 
continuous line. }  
\end{figure}

\begin{figure}
\includegraphics[width=0.70\textwidth,angle=90,scale=.75]{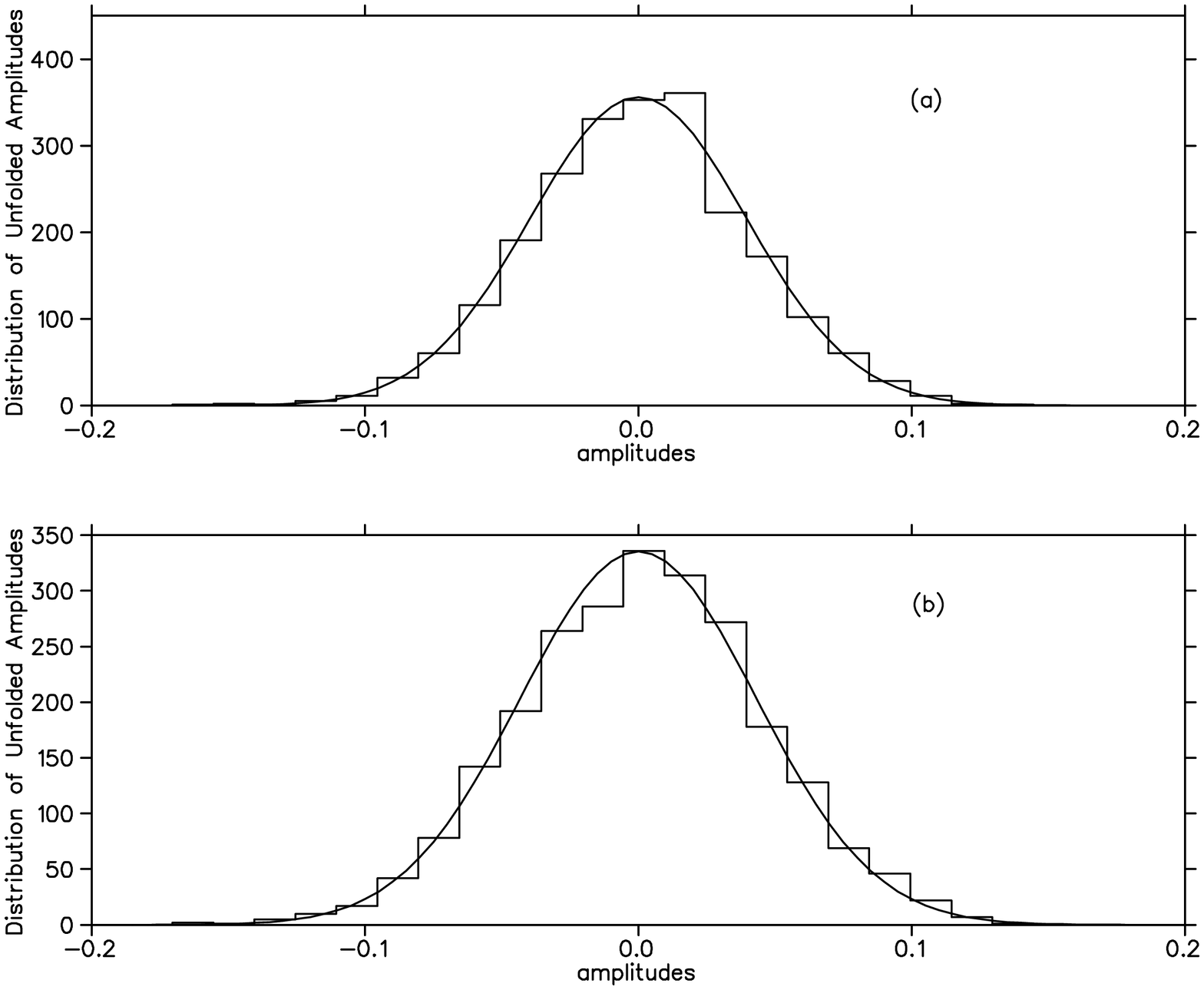}
\caption{\label{tailgaus} Histogram of the amplitudes distribution in a 
window of width of 20 energy units for 100 EFs around (a)$\psi_{900}$ of $H_1$ 
and (b)$\psi_{500}$ 
of $H_2$. The normalization corresponds to the number of events per bin, 
the total number of events being 2525 and 2595 for (a) and (b), respectively. 
The corresponding values of $\chi^2$ are  $\chi_{16}^2=12.289$ and
$\chi_{16}^2=14.483$. The continuous line corresponds
to a Gaussian curve with the same moments as the histograms.
}  
\end{figure}

We now test our assumption that scar effects are not seen when we
expand the chaotic Hamiltonian in a basis of another chaotic Hamiltonian,
instead of an integrable one. For such an expansion we put the EF of the $H_2$ 
Hamiltonian  in terms of
the $H_1$ EFs. The quantum-classical 
correspondence is shown in figure ~\ref{efsh2}.  In this case the exponential
decay in the classically forbidden area shows a hump, for which we have no 
explanation.
The B  zone is wider and in consequence the statistics are better as we show 
below. 
The amplitude 
distribution in the same region as in the previous case fits the Gaussian 
better,  
as shown in figure ~\ref{h1h2dist}. The excess of small amplitudes decreases
and the agreement is better in a wider energy regime. A similar result 
is observed if we drop localized states in the statistics for the previous
case. 
Beyond all these features, if we consider small windows in the tail of 
eigenfunctions we find statistically good Gaussians for both cases. 
In figure ~\ref{tailgaus}
we show some of them. The window width is of 20 mean level spacings  
in order to have a sufficient number of amplitudes ($\sim 1000$) of the 101 EFs
considered for the average. 
They have energies between 1000 and 1020 in figure~\ref{tailgaus}(a) for the
state 900 of $H_1$ and from 640 to 660 for the state 500 of $H_2$ in 
figure~\ref{tailgaus}(b). 
The fluctuations in figure~\ref{tailgaus} are larger than in the previous 
figures, but all of them are inside the statistical deviation, as shown by 
using the $\chi^2$ test per bin, which is $\chi_{16}^2=12.289$ for (a) and
$\chi_{16}^2=14.483$ for (b). For clarity we plot the histograms with
larger bins and normalized to (total number of events)$\times$(bin width). 
We cannot get such a good fit to the Gaussian distribution for all energy 
ranges; the larger the window width, the worse the observed fit.

\section{Conclusion}
We have analyzed the fluctuations of quantum-mechanical eigenfunctions with 
respect to their classical limit. Using a simple random-matrix model, 
the amplitudes are shown to follow a Gaussian distribution. This is confirmed
by a numerical calculation using systems of two particles interacting 
 through anharmonic potentials; agreement improves as we move up in the 
spectrum. We further find evidence for scars in 
an excess of small amplitude values as compared to the theoretical prediction 
if we express the eigenstate of the  chaotic Hamiltonian in term of an 
integrable one.
This effect decreases markedly when both Hamiltonians have chaotic dynamics.

\ack
This work was partially supported by the
DGAPA(UNAM) 
project IN-109201 and the
CONACyT(M\'exico) Grants No. 25192-E, 346668-E and 33773-E.
One of the authors(HHS) acknowledges the financial support by 
DGEP(UNAM) and PAEP-PCF(UNAM).

\newpage


\section*{References}
\begin{thebibliography}{99}
\bibitem{qch}
Berry M V and Tabor M 1977 {\it Proc. R. Soc. London} A {\bf 356} 375; 
Casati G, Guarneri I and Valz-Gris F 1980 {\it Lett. Nuovo 
Cimento} {\bf 28} 279;
Bohigas O,  Giannoni M-J and Schmit C 1983 {\it Phys. Rev. Lett.}
{\bf 52} 1;
Berry M V 1985 {\it Proc. R. Soc. London} A {\bf 400} 229;
Leyvraz F and Seligman T H 1992 {\it Phys. Lett.} A {\bf 168} 348.

\bibitem{rmt}
Mehta M L 1990 {\it Random Matrix Theory and Statistical Theory of
 Energy Levels} ( New York: Academic Press); Brody T A, Flores J,
French J B, Mello P A, Pandey A and Wong S S M 1981 {\it Rev. Mod. Phys.} 
{\bf 53} 385;
Guhr T, M\"uller-Groeling A and Weidenm\"uller H A 1998
{\it Phys. Rep.} {\bf 299} 189.
\bibitem{griba} Flambaum V V, Gribakina A A, Gribakin G F  and
Kozlov M G 1994 {\it Phys. Rev.} A {\bf 50} 267 ().
\bibitem{luna} Luna-Acosta  G A, M\'endez-Berm\'udez J A and Izrailev F M 2001
{\it Phys. Rev.} E {\bf 64} 036206.

\bibitem{2body} Benet L, Izrailev F M, Seligman T H and Suarez-Moreno A 2000 
{\it Phys. Lett.} A {\bf277} 87.
\bibitem{heller} Heller E J 1984 {\it Phys. Rev. Lett.} {\bf 53 } 1515.
\bibitem{kaplan} Kaplan L1999 {\it Nonlinearity} {\bf 12} R1. See  
references therein. See also Sridhar S, Lu W T 2002 {\it J. Stat Phys.} 
{\bf 108}
755. Wisniacki D A, Borondo F, Vergini E and Benito R M (2001) 
{it Phys. Rev.} E {\bf 65} 016213. 
\bibitem{thoulessenergy} Benvenuto F, Casati G, Shepeliansky D L 1997 
{\it Phys. Rev} A {\bf 55 } 1732.
\bibitem{beref} Berry M V 1977 {\it J. Phys. A: Math. Gen} {\bf 10} 2083.
\bibitem{hhst}Benet L, Flores J, Hern\'andez-Salda\~na H, Izrailev F M,
Leyvraz F and Seligman T H {\it to appear.}
\end{thebibliography}
\end{document}